\documentclass[article,12pt,superscriptaddress,showkeys,showpacs] {revtex4}
\usepackage{amsfonts,graphicx,amsmath,longtable}
\begin{document}
\title{Steady-state mechanical squeezing in a hybrid atom-optomechanical system with a highly dissipative cavity}
\author{Dong-Yang Wang}
\affiliation{Department of Physics, College of Science, Yanbian University, Yanji, Jilin 133002, People's Republic of China}
\author{Cheng-Hua Bai}
\affiliation{Department of Physics, College of Science, Yanbian University, Yanji, Jilin 133002, People's Republic of China}
\author{Hong-Fu Wang\footnote{E-mail: hfwang@ybu.edu.cn}}
\affiliation{Department of Physics, College of Science, Yanbian University, Yanji, Jilin 133002, People's Republic of China}
\affiliation{School of Physics, Northeast Normal University, Changchun, Jilin 130024, People's Republic of China}
\author{Ai-Dong Zhu}
\affiliation{Department of
Physics, College of Science, Yanbian University, Yanji, Jilin
133002, People's Republic of China}
\author{Shou Zhang\footnote{E-mail: szhang@ybu.edu.cn}}
\affiliation{Department of
Physics, College of Science, Yanbian University, Yanji, Jilin
133002, People's Republic of China}

\begin{abstract}
Quantum squeezing of mechanical resonator is important for studying the macroscopic quantum effects and the precision metrology of weak forces. Here we give a theoretical study of a hybrid atom-optomechanical system in which the steady-state squeezing of the mechanical resonator can be generated via the mechanical nonlinearity and cavity cooling process. The validity of the scheme is assessed by simulating the steady-state variance of the mechanical displacement quadrature numerically. The scheme is robust against dissipation of the optical cavity, and the steady-state squeezing can be effectively generated in a highly dissipative cavity.
\pacs{42.50.Pq, 07.10.Cm, 42.50.Lc, 42.50.Wk}
\keywords{mechanical squeezing, optomechanical system, dissipation}
\end{abstract}
\maketitle
\section{Introduction}
The optomechanical system is a rapidly growing field from the classical Fabry-P\'{e}rot interferometer by replacing one of the fixed sidewalls with a movable one~\cite{TKSCI08321}. The introduced one-dimensional freedom of the movable sidewall can be regarded as a free resonator mode, which can interact with the cavity mode through radiation pressure force originating from the light carrying momentum. Many projects of cavity optomechanics systems have been conceived and experimentally demonstrated in the past decade~\cite{MTFRMP1486,PAP13525,FSPHY092}. For example, the radiation force has been used for cooling the mechanical resonators to near their quantum ground states and entangling the cavity and mechanical resonator, and for coherent-state transiting between the cavity and mechanical resonator~\cite{JTDJNAT11475,ZTMPRA1183,XYPYPRA1592,YYCYCPB1322,TJRKSCI13342,YKWYPRA1490,
TRHPRA1592,VYMRPRL11107,HZYCPRA0878,MCMCPRA1388}. Quantum fluctuations become the dominant mechanical driving force with strong radiation pressure, which leads to correlations between the mechanical motion and the quantum fluctuations of the cavity field~\cite{TRCSCI13339}. In addition, the optomechanical method of manipulating the quantum fluctuations has also been used for generating the squeezing states of the optical and mechanical modes~\cite{TPRNPRX133,AFAPRA1388,JCPRA1183,XJLFPRA1591}.

The history of optical squeezing is linked intimately to quantum-limited displacement sensing~\cite{VF92}, and many schemes have been proposed to generate squeezing states in various systems~\cite{SPASPRA1388,RLBJPRL8555,MKGLNP084}. The squeezing of light field is proposed for the first time using atomic sodium as a nonlinear medium~\cite{RLBJPRL8555}. In addition, the squeezing of microwave field, which has been demonstrated with up to 10 dB of noise suppression~\cite{MKGLNP084}, is an important tool in quantum information processing with superconducting circuits. In recent years, researchers have found that the optomechanical cavity, which can be regard as a low-noise Kerr nonlinear medium~\cite{CMSAPRA9449,SPPRA9449}, can be a better candidate to generate squeezing of the optical and mechanical modes. The squeezing of optical field is easy to be achieved in the optomechanical systems, and has been obtained experimentally~\cite{TPRNPRX133,ASJJNAT13500,DTSTNAT12488}. However, the squeezing of mechanical mode has not been observed experimentally. Many schemes have been proposed to generate mechanical squeezing in the optomechanical systems, including methods based on measurement, feedback, parametric processes, and the concept of quantum-reservoir engineering~\cite{AFKNJP0810,AJPRL09103,WGYPRA1388,JYFPRA0979,MMPB00280,PAPPRB0470}. Quantum squeezing of mechanical mode is one of the key macroscopic quantum effects, which can be used for studying the quantum-to-classical transition and improving the precision of quantum measurements~\cite{RLBJPRL8555,LHJHPRL8657,CKRVRMP8052,XFPRL9676}. So the mechanical squeezing attracts more and more attentions. For example, in 2011, Liao $et~al.$~\cite{JCPRA1183} proposed a scheme to generate mechanical squeezing in a optomechanical cavity. They showed that parametric resonance could be reached approximately by periodically modulating  the driving field amplitude at a frequency matching the frequency shift of the mirror, leading to an efficient generation of squeezing. In 2013, Kronwald $et~al.$~\cite{AFAPRA1388} proposed a scheme to generate mechanical squeezing by driving the optomechanical cavity with two controllable lasers with differing amplitudes. The scheme utilized a dissipative mechanism with the driven cavity acting as an engineered reservoir. In 2015, L\"{u} $et~al.$~\cite{XJLFPRA1591} proposed a scheme to generate steady-state mechanical squeezing via mechanical nonlinearity, which showed that squeezing could be achieved by the joint effect of nonlinearity-induced parametric amplification and cavity cooling process.

Traditionally and generally, the decay rate of cavity field, which is a dissipative factor in optomechanical system, is considered to have negative effect on the performance of quantum manipulation of mechanical modes. Here we propose a method to generate steady-state mechanical squeezing in a hybrid atom-optomechanical system where the atomic ensemble is trapped in the optical cavity consisting of a fixed mirror and a movable mirror. The coherently driving on the cavity mode is a monochromatic laser source which can generate strong optomechanical coupling between the mechanical and cavity modes. We show that, via the mechanical nonlinearity and cavity cooling process in transformed frame, the steady-state mechanical squeezing can be successfully and effectively generated in a highly dissipative cavity.

The paper is organized as follows: In Section \uppercase\expandafter{\romannumeral2}, we describe the model of a hybrid atom-optomechanical system and derive the effective coupling between the atomic ensemble and the mechanical resonator. In Section \uppercase\expandafter{\romannumeral3}, we engineer the mechanical squeezing and derive the analytical variance of the displacement quadrature of the movable mirror in the steady-state. In Section \uppercase\expandafter{\romannumeral4}, we study the variance of mechanical mode with the large decay rate of cavity by numerical simulations method and discuss the validity of the scheme in the highly and lowly dissipative cavities. A conclusion is given in Section \uppercase\expandafter{\romannumeral5}.
\section{System and Model}
\begin{figure}
  \includegraphics[width=3in]{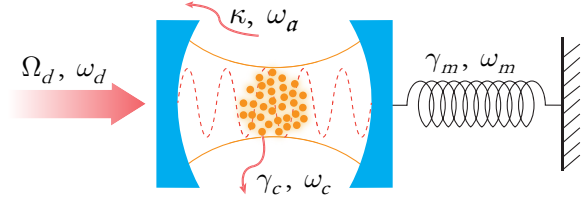}\\
  \caption{(Color online) Schematic diagram of a hybrid atom-optomechanical system with a cloud of identical two-level atoms trapped in an optical cavity consisting of a fixed mirror and a movable mirror. The cavity mode is coherently driven by an input laser with frequency $\omega_{d}$.}\label{f01}
\end{figure}

We consider a hybrid atom-optomechanical system depicted in Fig.~\ref{f01}, in which $N$ identical two-level atoms are trapped in the optical cavity consisting of a fixed mirror and a movable mirror. The total Hamiltonian $H=H_{0}+H_{\mathrm{I}}+H_{\mathrm{pump}}$, which describes the hybrid system, consists of three parts, which reads ($\hbar=1$), respectively,
\begin{eqnarray}\label{e01}
  H_{0}&=&\omega_{a}a^{\dag}a+\omega_{c}S_{z}
        +\omega_{m}b^{\dag}b+\eta\left(b+b^{\dag}\right)^{3},\cr\cr
  H_{\mathrm{I}}&=&\bar{g}_{0}\left(S_{-}a^{\dag}+S_{+}a\right)-ga^{\dag}a\left(b+b^{\dag}\right),\cr\cr
  H_{\mathrm{pump}}&=&\Omega_{d}\left(e^{-i\omega_{d}t}a^{\dag}+e^{i\omega_{d}t}a\right).
\end{eqnarray}
The part $H_{0}$ accounts for the free Hamiltonian of the cavity mode (with frequency $\omega_{a}$ and decay rate $\kappa$), the atoms (with transition frequency $\omega_{c}$ and linewidth $\gamma_{c}$), and the mechanical resonator (with frequency $\omega_{m}$ and damping rate $\gamma_{m}$). Here $a~(a^{\dag})$ is the bosonic annihilation (creation) operator of the optical cavity mode, $b~(b^{\dag})$ is the bosonic annihilation (creation) operator of the mechanical mode, and $S_{\mathrm{z}}=\sum_{i=1}^{N}\sigma_{z}^{i}$ is the collective $z-$spin operator of the atoms. The last term of $H_{0}$ describes the cubic nonlinearity of the mechanical resonator with amplitude $\eta$. For mechanical resonator in the gigahertz range, the intrinsic nonlinearity is usually very weak with nonlinear amplitude smaller than $10^{-15}\omega_{m}$. We can obtain a strong nonlinearity through coupling the mechanical mode to an ancillary system~\cite{MPR04395,ZSJFRMP1385,KAPRL09103,LPRB1184}, such as the nonlinear amplitude of $\eta=10^{-4}\omega_{m}$ can be obtained when we couple the mechanical resonator to an external qubit~\cite{XJLFPRA1591}.

The part $H_{\mathrm{I}}$ accounts for the interaction Hamiltonian consisting of the atom-field interaction and the optomechanical interaction derived from the radiation pressures. Where $\bar{g}_{0}=\sum_{i=1}^{N}g_{0}^{i}/N$ represents the averaged atom-field coupling strength with $g_{0}^{i}$ being the coupling strength between the $i$th atom and single-photon, and $g$ is the single-photon optomechanical coupling strength.

The part $H_{\mathrm{pump}}$ accounts for the external driving laser with frequency $\omega_{d}$ used to coherently pump the cavity mode. The driving strength $\Omega_{\mathrm{d}}=\sqrt{2P\kappa/(\hbar\omega_{d}})$ is related to the input laser power $P$, the mechanical resonator frequency $\omega_{d}$, and the decay rate of cavity $\kappa$.

\begin{figure}
  \includegraphics[width=3.5in]{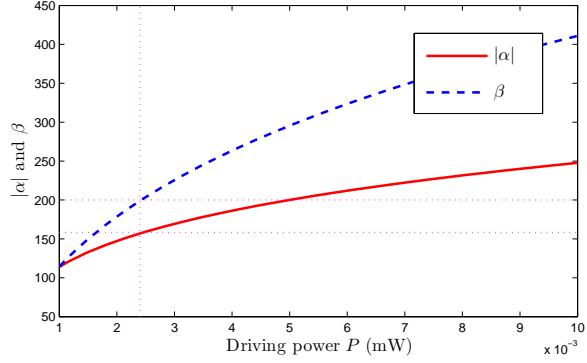}\\
  \caption{(Color online) The steady-state amplitudes $|\alpha|$ and $\beta$ versus the driving power $P$. The parameters are chosen to be $\omega_{m}/(2\pi)=5~\mathrm{MHz}$, $\omega_{a}/(2\pi)=500~\mathrm{THz}$, $\delta_{a}=2\omega_{m}$, $\Delta_{c}=\omega_{m}$, $G_{0}=0.5\omega_{m}$, $g=10^{-2}\omega_{m}$, $\eta=10^{-4}\omega_{m}$, $\kappa=10\omega_{m}$, $\gamma_{c}=0.1\omega_{m}$, $\gamma_{m}=10^{-6}\omega_{m}$, and $\Omega_{d}=\sqrt{2P\kappa/(\hbar\omega_{d}})$.}\label{f02}
\end{figure}

The spin operators $S_{-}~(S_{+})$ of the atomic ensemble can be transformed to a collective bosonic operator $c~(c^{\dag})$ in the Holstein-Primakoff representation~\cite{XYPYPRA1592,SPASPRA1388},
\begin{eqnarray}\label{e02}
  S_{-}&=&c\sqrt{N-c^{\dag}c}\simeq\sqrt{N}c,\cr\cr
  S_{+}&=&c^{\dag}\sqrt{N-c^{\dag}c}\simeq\sqrt{N}c^{\dag},\cr\cr
  S_{\mathrm{z}}&=&c^{\dag}c-\frac{N}{2},
\end{eqnarray}
where operators $c$ and $c^{\dag}$ obey the standard boson commutator $[c,~c^{\dag}]=1$. Under the conditions of sufficiently large atom number $N$ and weak atom-photon coupling $\bar{g}_{0}$, the total Hamiltonian in the frame rotating at input laser frequency $\omega_{d}$ is written as
\begin{eqnarray}\label{e03}
  H^{'}&=&-\delta_{a}a^{\dag}a-\Delta_{c}c^{\dag}c+\omega_{m}b^{\dag}b
        +\eta\left(b+b^{\dag}\right)^{3}+G_{0}\left(a^{\dag}c+ac^{\dag}\right)\cr\cr
        &&-ga^{\dag}a\left(b+b^{\dag}\right)+\Omega_{d}\left(a+a^{\dag}\right),
\end{eqnarray}
where $\delta_{a}=\omega_{d}-\omega_{a}$, $\Delta_{c}=\omega_{d}-\omega_{c}$, and $G_{0}=\bar{g}_{0}\sqrt{N}$. Applying a displacement transformation to linearize the Hamiltonian, $a\rightarrow\alpha+a,~b\rightarrow\beta+b,~c\rightarrow\xi+c$, where $\alpha,~\beta$, and $\xi$ are $c$ numbers denoting the steady-state amplitude of the cavity, mechanical, and collective atomic modes, which are derived by solving the following equations:
\begin{eqnarray}\label{e04}
\left[i\left(\delta_{a}+2g\beta\right)-\frac{\kappa}{2}\right]\alpha-iG_{0}\xi-i\Omega_{d}=0,\cr\cr
\omega_{m}\beta+3\eta\left(4\beta^{2}+1\right)-g\left|\alpha\right|^2=0,\cr\cr
\left(i\Delta_{c}-\frac{\gamma_{c}}{2}\right)\xi-iG_{0}\alpha=0.
\end{eqnarray}
Under the conditions of $\gamma_{m}\ll\kappa,~\gamma_{c},~\eta$, the $\gamma_{m}$-dependent terms can be neglected. One can see that when the driving power $P$ is in the microwatt range, the amplitudes of the cavity and mechanical modes satisfy the relationships: $|\alpha|,~\beta\gg1$, as shown in Fig.~\ref{f02}. And the amplitudes of the cavity and mechanical modes increase with increasing the driving power. For example, at the point of the driving power $P=2.4\times10^{-3}~\mathrm{mW}$, $|\alpha|\simeq160$ and $\beta\simeq200$ can be obtained, respectively.

After the standard linearization procedure, the linearized Hamiltonian is given by
\begin{eqnarray}\label{e05}
  H_{\mathrm{L}}=-\Delta_{a}a^{\dag}a-\Delta_{c}c^{\dag}c+\tilde{\omega}_{m}b^{\dag}b
        +\Lambda\left(b^{2}+b^{\dag2}\right)+G_{0}\left(a^{\dag}c+ac^{\dag}\right)
        -G\left(a+a^{\dag}\right)\left(b+b^{\dag}\right),
\end{eqnarray}
with
\begin{eqnarray}\label{e06}
    \Delta_{a}&=&\delta_{a}+2g\beta,~~~\tilde{\omega}_{m}=\omega_{m}+2\Lambda,\cr\cr
    \Lambda&=&6\eta\beta,~~~~~~~~~~~G=g|\alpha|.
\end{eqnarray}
And the Hamiltonian of the nonlinear terms, which come from the radiation-pressure interaction and the cubic nonlinearity, is written as
\begin{eqnarray}\label{e07}
  H_{\mathrm{NL}}=-ga^{\dag}a\left(b+b^{\dag}\right)+\left(3\eta b^{\dag2}b+\eta b^{\dag3}+\mathrm{H.c.}\right).
\end{eqnarray}
Under the conditions of $g,~\eta\ll\Lambda,~G,~G_{0}$, the nonlinear terms in $H_{\mathrm{NL}}$ can be neglected because they are much weaker than the linear terms in $H_{\mathrm{L}}$.

Considering the effect of the thermal environment and basing on the linearized Hamiltonian $H_{\mathrm{L}}$, the quantum Langevin equations for the system are written as
\begin{eqnarray}\label{e08}
    \dot{a}&=&\left(i\Delta_{a}-\frac{\kappa}{2}\right)a-iG_{0}c+iG\left(b+b^{\dag}\right)
            -\sqrt{\kappa}a_{in},\cr\cr
    \dot{b}&=&\left(-i\tilde{\omega}_{m}-\frac{\gamma_{m}}{2}\right)b+iG\left(a+a^{\dag}\right)
    -2i\Lambda b^{\dag}-\sqrt{\gamma_{m}}b_{in},\cr\cr
    \dot{c}&=&\left(i\Delta_{c}-\frac{\gamma_{c}}{2}\right)c-iG_{0}a-\sqrt{\gamma_{c}}c_{in},
\end{eqnarray}
where the corresponding noise operators $a_{in},~b_{in}$, and $c_{in}$ satisfy correlations $\langle a_{in}(t)a_{in}^{\dag}(t^{'})\rangle=\langle c_{in}(t)c_{in}^{\dag}(t^{'})\rangle=\delta(t-t^{'}),~\langle a_{in}^{\dag}(t)a_{in}(t^{'})\rangle=\langle c_{in}^{\dag}(t)c_{in}(t^{'})\rangle=0,~\langle b_{in}(t)b_{in}^{\dag}(t^{'})\rangle=(\bar{n}_{\mathrm{th}}+1)\delta(t-t^{'}),~\langle b_{in}^{\dag}(t)b_{in}(t^{'})\rangle=\bar{n}_{\mathrm{th}}\delta(t-t^{'})$, where $\bar{n}_{\mathrm{th}}=\left\{\mathrm{exp}\left[\hbar\omega_{m}/(k_{B}T)\right]-1\right\}^{-1}$ is the mean thermal excitation number of bath of the movable mirror at temperature $T$, $k_{B}$ is the Boltzmann constant, and one recovers a Markovian process. Since the decay rate of cavity, $\kappa$, is much larger than the linewidth of the atoms, and under the conditions $|\Delta_{a}|\gg|\Delta_{c}|,~\tilde{\omega}_{m}\gg2\Lambda,~\kappa\gg(\gamma_{c},
~\omega_{m}),~\omega_{m}\gg\gamma_{m}$, we can approximatively obtain~\cite{XYPYPRA1592}
\begin{eqnarray}\label{e09}
    a(t)\simeq\frac{iG[b(t)+b^{\dag}(t)]}{-i\Delta_{a}+\frac{\kappa}{2}}
        -\frac{iG_{0}c(t)}{-i\Delta_{a}+\frac{\kappa}{2}}
        +a(0)\mathrm{exp}\left(i\Delta_{a}t-\frac{\kappa}{2}t\right)+A_{in}^{'}(t),
\end{eqnarray}
where $A_{in}^{'}(t)$ denotes the noise term. Neglecting the fast decaying term which contains $\mathrm{exp}(-\kappa t/2)$ and substituting Eq.~(\ref{e09}) into Eq.~(\ref{e08}), we can obtain the effective coupling between the mechanical mode $b$ and collective atoms mode $c$, which can be written as
\begin{eqnarray}\label{e10}
    \dot{b}&=&\left(-i\tilde{\omega}_{m}^{'}-\frac{\gamma_{m}}{2}\right)b
    +iG_{\mathrm{eff}}\left(c+c^{\dag}\right)-2i\Lambda^{'}b^{\dag}-\sqrt{\gamma_{m}}b_{in},\cr\cr
    \dot{c}&=&\left(i\Delta_{\mathrm{eff}}-\frac{\gamma_{\mathrm{eff}}}{2}\right)c
            +iG_{\mathrm{eff}}\left(b+b^{\dag}\right)-\sqrt{\gamma_{c}}c_{in},
\end{eqnarray}
where the effective parameters of the mechanical frequency, optomechanical coupling strength, detuning, damping rate, and coefficients of bilinear terms are given by
\begin{eqnarray}\label{e11}
  \tilde{\omega}_{m}^{'}&=&\tilde{\omega}_{m}
  +\frac{2G^{2}\Delta_{a}}{\Delta_{a}^{2}+\left(\frac{\kappa}{2}\right)^{2}},\cr\cr
  G_{\mathrm{eff}}&=&\left|\frac{GG_{0}}{\Delta_{a}+i\frac{\kappa}{2}}\right|,\cr\cr
  \Delta_{\mathrm{eff}}&=&\Delta_{c}-\frac{G_{0}^{2}\Delta_{a}}{\Delta_{a}^{2}
  +\left(\frac{\kappa}{2}\right)^{2}},\cr\cr
  \gamma_{\mathrm{eff}}&=&\gamma_{c}+\frac{G_{0}^{2}\kappa}{\Delta_{a}^{2}
  +\left(\frac{\kappa}{2}\right)^{2}},\cr\cr
  \Lambda^{'}&=&\Lambda+\frac{G^{2}\Delta_{a}}{\Delta_{a}^{2}+\left(\frac{\kappa}{2}\right)^{2}}.
\end{eqnarray}
Thus the effective Hamiltonian is rewritten as
\begin{eqnarray}\label{e12}
    H_{\mathrm{eff}}=-\Delta_{\mathrm{eff}}c^{\dag}c+\tilde{\omega}_{m}^{'}b^{\dag}b
            -G_{\mathrm{eff}}\left(c+c^{\dag}\right)\left(b+b^{\dag}\right)
            +\Lambda^{'}\left(b^{\dag2}+b^{2}\right).
\end{eqnarray}

When considering the system-reservoir interaction, which results in the dissipations of the system, the full dynamics of the effective system is described by the master equation
\begin{eqnarray}\label{e13}
\dot{\rho}=-i\left[H_{\mathrm{eff}},\rho\right]+\gamma_{\mathrm{eff}}\mathcal{L}[c]\rho
    +\gamma_{m}\left(\bar{n}_{\mathrm{th}}+1\right)\mathcal{L}[b]\rho
    +\gamma_{m}\bar{n}_{\mathrm{th}}\mathcal{L}[b^{\dag}]\rho,
\end{eqnarray}
where $\mathcal{L}[o]\rho=o\rho o^{\dag}-(o^{\dag}o\rho+\rho o^{\dag}o)/2$ is the standard Lindblad operators, $\gamma_{\mathrm{eff}}$ is the effective damping rate of the mode $c$, and $\bar{n}_{\mathrm{th}}$ is the average phonon number in thermal equilibrium.
\section{Engineering the mechanical squeezing}
Applying the unitary transformation $S(\zeta)=\mathrm{exp}[\zeta(b^{2}+b^{\dag2})/2]$, which is the single-mode squeezing operator with the squeezing parameter
\begin{eqnarray}\label{e14}
    \zeta=\frac{1}{4}\mathrm{ln}\left(1+\frac{4\Lambda^{'}}{\omega_{m}}\right),
\end{eqnarray}
to the total system. Then the transformed effective Hamiltonian becomes
\begin{eqnarray}\label{e15}
    H_{\mathrm{eff}}^{'}=S^{\dag}\left(\zeta\right)H_{\mathrm{eff}}S\left(\zeta\right)=
    -\Delta_{\mathrm{eff}}c^{\dag}c+\omega_{m}^{'}b^{\dag}b
    -G^{'}\left(c+c^{\dag}\right)\left(b+b^{\dag}\right),
\end{eqnarray}
with
\begin{eqnarray}\label{e16}
\omega_{\mathrm{m}}^{'}&=&\omega_{m}\sqrt{1+\frac{4\Lambda^{'}}{\omega_{m}}},\cr\cr G^{'}&=&G_{\mathrm{eff}}\left(1+\frac{4\Lambda^{'}}{\omega_{m}}\right)^{-\frac{1}{4}},
\end{eqnarray}
where $\omega_{\mathrm{m}}^{'}$ is the transformed effective mechanical frequency and $G^{'}$ is the transformed effective optomechanical coupling. The transformed Hamiltonian is a standard cavity cooling Hamiltonian and the best cooling in the transformed system is at the optimal detuning $\Delta_{\mathrm{eff}}=-\omega_{m}^{'}$. In the transformed frame, the master equation $\rho^{'}=S^{\dag}(\zeta)\rho S(\zeta)$ of system-reservoir interaction can be approximatively written as~\cite{XJLFPRA1591}
\begin{eqnarray}\label{e17}
\dot{\rho}^{'}&=&-i\left[H_{\mathrm{eff}}^{'},\rho^{'}\right]+\gamma_{\mathrm{eff}}\mathcal{L}[c]\rho^{'}
    +\gamma_{m}\left(\bar{n}_{\mathrm{th}}^{'}+1\right)\left(\mathrm{cosh}^{2}(\zeta)\mathcal{L}[b]
    +\mathrm{sinh}^{2}(\zeta)\mathcal{L}[b^{\dag}]\right)\rho^{'}\cr\cr
    &&+\gamma_{m}\bar{n}_{\mathrm{th}}^{'}\left(\mathrm{cosh}^{2}(\zeta)\mathcal{L}[b^{\dag}]
    +\mathrm{sinh}^{2}(\zeta)\mathcal{L}[b]\right)\rho^{'},
\end{eqnarray}
which is the transformed master equation and can achieve the cooling process. Here $\bar{n}_{\mathrm{th}}^{'}=\bar{n}_{\mathrm{th}}\mathrm{cosh}(2\zeta)
+\mathrm{sinh}^{2}(\zeta)$, is the transformed thermal phonon number. The steady-state density matrix $\rho$ can be obtained by solving the master equation Eq.~(\ref{e13}) numerically. Defining the displacement quadrature $X=b+b^{\dag}$ for the mechanical mode, the steady-state variance of $X$ is given by $\langle\delta X^{2}\rangle=\langle X^{2}\rangle-\langle X\rangle^{2}$, which can be derived as
\begin{eqnarray}\label{e18}
    \langle\delta X^{2}\rangle=\left(2\bar{n}_{\mathrm{eff}}^{'}+1\right)e^{-2\zeta},
\end{eqnarray}
where $\bar{n}_{\mathrm{eff}}^{'}$ is the steady-state phonon number of the transformed system. When the best cooling (at the optimal detuning $\Delta_{\mathrm{eff}}=-\omega_{m}^{'}=-\omega_{m}\sqrt{1+4\Lambda^{'}/\omega_{m}}$) in the transformed system $\bar{n}_{\mathrm{eff}}^{'}=0$ is achieved by the cooling process, the steady-state variance $\langle\delta X^{2}\rangle=e^{-2\zeta}$ approaches the minimum value.

\section{Numerical simulations and discussion}
\begin{figure}
  \includegraphics[width=3.5in]{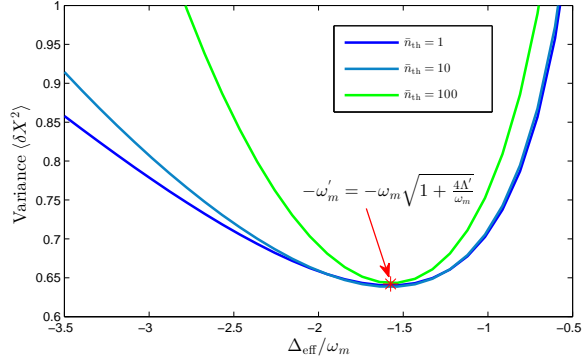}\\
  \caption{(Color online) The variance of the displacement quadrature $X$ relates to the effective detuning $\Delta_{\mathrm{eff}}$ by solving the master equation numerically. Here $\Delta_{\mathrm{eff}}$ can be tuned individually by varying $\Delta_{c}$, the average phonon number $\bar{n}_{\mathrm{th}}$ is set to be $1$, $10$, and $100$ respectively, and the other parameters are chosen to be the same as in Fig.~\ref{f02}.}\label{f03}
\end{figure}

In this section, we solve the master equation Eq.~(\ref{e13}) numerically to calculate the steady-state variance of the mechanical displacement quadrature $X$. The relationship between the steady-state variance and effective detuning is shown in Fig.~\ref{f03}. One can see from Fig.~\ref{f03} that the minimum value of variance can be achieved at the optimal detuning point of $\Delta_{\mathrm{eff}}=-\omega_{m}^{'}$, which comes from the standard cavity cooling Hamiltonian in Eq.~(\ref{e15}) under the transformed frame. The change rate of variance on the effective detuning increases with increasing the average phonon number $\bar{n}_{\mathrm{th}}$. In the process of numerical simulation, the parameters are set to be $\omega_{m}/(2\pi)=5~\mathrm{MHz}$, $\omega_{a}/(2\pi)=500~\mathrm{THz}$, $\delta_{a}=2\omega_{m}$, $\Delta_{c}=\omega_{m}$, $G_{0}=0.5\omega_{m}$, $g=10^{-2}\omega_{m}$, $\eta=10^{-4}\omega_{m}$, $\kappa=10\omega_{m}$, $\gamma_{c}=0.1\omega_{m}$, $\gamma_{m}=10^{-6}\omega_{m}$, $\Omega_{d}=\sqrt{2P\kappa/(\hbar\omega_{d}})$, and $\bar{n}_{\mathrm{th}}=1,~10,~100$ respectively, which satisfy the conditions $|\Delta_{a}|\gg|\Delta_{c}|,~\tilde{\omega}_{m}\gg2\Lambda,~\kappa\gg(\gamma_{c},
~\omega_{m}),~\omega_{m}\gg\gamma_{m},~(\kappa,~\gamma_{c},~\eta)\gg\gamma_{m},~(|\alpha|,~\beta)\gg1$, and $(\Lambda,~G,~G_{0})\gg(g,~\eta)$. The average phonon number $\bar{n}_{\mathrm{th}}=100$ corresponds to the temperature $T=25\mathrm{mK}$. At the optimal detuning point $\Delta_{\mathrm{eff}}=-\omega_{m}^{'}=-\omega_{m}\sqrt{1+4\Lambda^{'}/\omega_{m}}$, the steady-state variance of the displacement quadrature is $\langle\delta X^{2}\rangle=e^{-2\zeta}=0.64$. However, one can see from Fig.~\ref{f03} that we need a more precise control for $\Delta_{\mathrm{eff}}$ to achieve the optimal steady-state squeezing of the mechanical resonator with the temperature rising constantly.

\begin{figure}
  \includegraphics[width=3.5in]{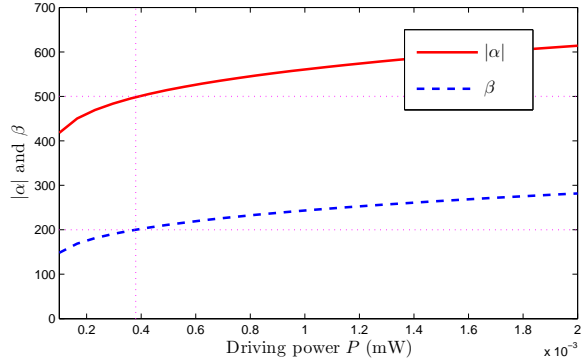}\\
  \caption{(Color online) The steady-state amplitudes $|\alpha|$ and $\beta$ versus the driving power $P$. The parameters are chosen to be $\omega_{m}/(2\pi)=5~\mathrm{MHz}$, $\omega_{a}/(2\pi)=500~\mathrm{THz}$, $\delta_{a}=-0.25\omega_{m}$, $\Delta_{c}=0.01\omega_{m}$, $G_{0}=0.05\omega_{m}$, $g=10^{-3}\omega_{m}$, $\eta=10^{-4}\omega_{m}$, $\kappa=0.1\omega_{m}$, $\gamma_{c}=0.1\omega_{m}$, $\gamma_{m}=10^{-6}\omega_{m}$, and $\Omega_{d}=\sqrt{2P\kappa/(\hbar\omega_{d}})$.}\label{f04}
\end{figure}

\begin{figure}
  \includegraphics[width=3.5in]{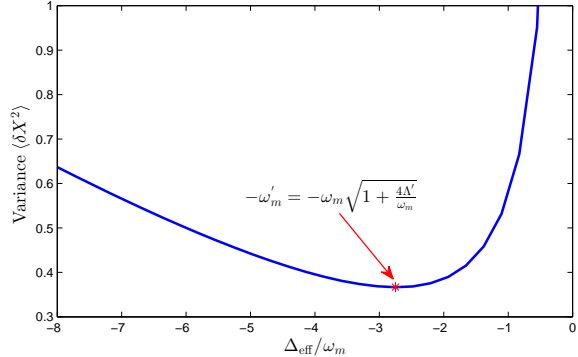}\\
  \caption{(Color online) The variance of the mechanical displacement quadrature $X$ relates to the effective detuning $\Delta_{\mathrm{eff}}$ by solving the master equation numerically. The average phonon number is set to $\bar{n}_{\mathrm{th}}=1$ and the other parameters are chosen to be the same as in Fig.~\ref{f04}.}\label{f05}
\end{figure}

In addition, considering the smaller decay rate of cavity, for example, $\kappa=0.1\omega_{m}$, and with the choices of $\delta_{a}=-0.25\omega_{m}$, $\Delta_{c}=0.01\omega_{m}$, $G_{0}=0.05\omega_{m}$, $g=10^{-3}\omega_{m}$, $\eta=10^{-4}\omega_{m}$, $\gamma_{c}=0.1\omega_{m}$, and $\gamma_{m}=10^{-6}\omega_{m}$. The relationship between the steady-state amplitudes $(|\alpha|,~\beta)$ and driving power $P$ is shown in Fig.~\ref{f04} and the relationship between the steady-state variance and effective detuning is shown in Fig.~\ref{f05} (here we calculate the steady-state variance of the mechanical displacement quadrature $X$ numerically by setting $P=0.38\times10^{-3}\mathrm{mW}$, $|\alpha|=500$, and $\beta=200$), respectively. At the optimal detuning point $\Delta_{\mathrm{eff}}=-\omega_{m}^{'}=-\omega_{m}\sqrt{1+4\Lambda^{'}/\omega_{m}}$, the steady-state variance of the displacement quadrature is $\langle\delta X^{2}\rangle=e^{-2\zeta}=0.36$.

In the above, we study the steady-state squeezing of the mechanical resonator in a hybrid atom-optomechanical system and illustrate that it can be effectively generated in both the highly and lowly dissipative cavities. The steady-state squeezing can be generated at the optimal detuning point by adjusting the parameters appropriately. When the decay rate of cavity is known, the maximum value of the squeezing parameter $\zeta$ is achieved at the point of $\Delta_{a}=\kappa/2$, which can be easily seen from Eq.~(\ref{e11}). Furthermore, the generated steady-state mechanical squeezing in the present scheme can be detected based on the method proposed in Refs.~\cite{XJLFPRA1591,DSAHPRL0798}. As illustrated in Refs.~\cite{XJLFPRA1591,DSAHPRL0798}, for detecting the steady-state mechanical squeezing, we can measure the position and the momentum quadratures of the mechanical resonator via homodyning detection of the output field of another auxiliary cavity mode with an appropriate phase, and the auxiliary cavity is driven by another pump laser field under a much weaker intracavity field so that its backaction on the mechanical mode can be neglected.

\section{Conclusions}
In conclusion, we have proposed a scheme for generating the steady-state squeezing of the mechanical resonator in a hybrid atom-optomechanical system via the mechanical nonlinearity and cavity cooling process in transformed frame. The atomic ensemble is trapped in the optomechanical cavity, which is driven by an external monochrome laser. The effective coupling between the mechanical resonator and the atomic ensemble can be obtained by reducing the cavity mode in the case of large detuning. We simulate the steady-state variance of the mechanical displacement quadrature numerically at a determinate laser driving power and find that the steady-state variance has the minimum value at the optimal detuning point, where the effective detuning is in resonance with the effective transformed mechanical frequency.

\begin{center}
{\bf{ACKNOWLEDGMENTS}}
\end{center}

This work was supported by the National Natural Science Foundation of China under
Grant Nos. 11264042, 11465020, 61465013, 11165015, and 11564041.

\end{document}